\documentclass[%
 reprint,
 amsmath,amssymb,
 aps,
prb,
]{revtex4-1}
\usepackage[utf8]{inputenc}
\usepackage{graphicx}
\usepackage{dcolumn}
\usepackage{bm}
\usepackage{hyperref}
\usepackage{amsmath}
\usepackage{amssymb}
\usepackage{pdfpages}
\makeatletter
\AtBeginDocument{\let\LS@rot\@undefined}
\makeatother

\begin{document}
\title{Electrostatically induced quantum point contact in bilayer graphene}
\author{Hiske Overweg}
\email{overwegh@phys.ethz.ch}
\author{Hannah Eggimann}
\affiliation{%
Solid State Physics Laboratory, ETH Zürich,~CH-8093~Zürich, Switzerland}
\author{Xi Chen}
\author{Sergey Slizovskiy}
\affiliation{National Graphene Institute, University of Manchester, Manchester M13 9PL, UK}
\altaffiliation{On leave of absence from NRC ``Kurchatov Institute'' PNPI, Russia}
\author{Marius Eich}

\author{Riccardo Pisoni}
\author{Yongjin Lee}
\author{Peter Rickhaus}
\affiliation{%
Solid State Physics Laboratory, ETH Zürich,~CH-8093~Zürich, Switzerland}

\author{Kenji Watanabe}
\author{Takashi Taniguchi}
\affiliation{National Institute for Material Science, 1-1 Namiki, Tsukuba 305-0044, Japan}

\author{Vladimir Fal'ko}
\affiliation{National Graphene Institute, University of Manchester, Manchester M13 9PL, UK}

\author{Thomas Ihn}
\author{Klaus Ensslin}

\affiliation{%
Solid State Physics Laboratory, ETH Zürich,~CH-8093~Zürich, Switzerland}

\begin{abstract}
We report the fabrication of electrostatically defined nanostructures in encapsulated bilayer graphene, with leakage resistances below depletion gates as high as $R \sim 10~$G$\Omega$. This exceeds previously reported values of $R =~$10 - 100 k$\Omega$.\cite{Zou2010,Yan2010,Zhu2016a} We attribute this improvement to the use of a graphite back gate. We realize two split gate devices which define an electronic channel on the scale of the Fermi-wavelength. A channel gate covering the gap between the split gates varies the charge carrier density in the channel. We observe device-dependent conductance quantization of $\Delta G = 2~e^2/h$ and $\Delta G = 4~e^2/h$. In quantizing magnetic fields normal to the sample plane, we recover the four- fold Landau level degeneracy of bilayer graphene. Unexpected mode crossings appear at the crossover between zero magnetic field and the quantum Hall regime.
\end{abstract}
\maketitle
Nanostructures in graphene offer unique perspectives in terms of confinement strength, device geometry and possible spin coherence. In single layer graphene the formation of tunnel barriers, a fundamental building block of any nanostructure, has been demonstrated by many experiments in which narrow channels were defined by dry etching. These experiments suffer from randomly positioned localized states along the sample edges.\cite{Bischoff2016,Bischoff2014} As a consequence, the barrier transmission cannot be tuned monotonically by electrostatic gates.\cite{Bischoff2016,Bischoff2014}. Bilayer graphene offers a promising alternative since a vertical electric field opens a band gap, which allows for depletion of the system. Several research groups used this property to define one-dimensional channels or quantum dots \cite{Allen2012,Goossens2012a,Droscher2012}, where the carriers are guided via a split gate structure with depleted graphene regions below the biased split gates. For the experiments published so far, the minimal conductance achievable in such geometries is limited by leakage currents below the split gates, presumably caused by hopping transport or a small energy gap. For tunnel barriers to be useful for high-quality quantum devices, the tunnelling resistance should exceed the resistance quantum $h/e^2$ by far\cite{Kouwenhoven1997}.

In this work we present two ultra-clean bilayer graphene samples encapsulated in hexagonal boron nitride (hBN) with a homogeneous top gate stripe crossing the current path in combination with a global graphite back gate. When depleting the region below the top gate, we measure resistances up to $10^5 \times~h/e^2$. In a next step, a split gate geometry was added to the devices, which was then covered by another insulating layer and a gate on top of the channel. In GaAs, similar QPC gate geometries have been studied\cite{Um2012}. This gate combination allows us to define an electron channel with resistances exceeding $1000 \times~h/e^2$ when depleted. The combination of top gates and back gate is essential to separately tune  the gap and the position of the Fermi level in the regions underneath the split gates as well as the carrier density in the channel.
When the channel gate voltage is increased above the depletion voltage, the electron channel is opened and the conductance displays plateaus. For sample $A$ the plateaus occur at conductance values 8, 10, 12, ..., 18 $e^2/h$ and for sample $B$ at 4, 8, 12 $e^2/h$.  With increasing magnetic field perpendicular to the two-dimensional layer, we observe mode mixing and mode crossing evolving into the expected Landau level spectrum for high magnetic fields.

\begin{figure}
\centering
\includegraphics[width=0.5\textwidth]{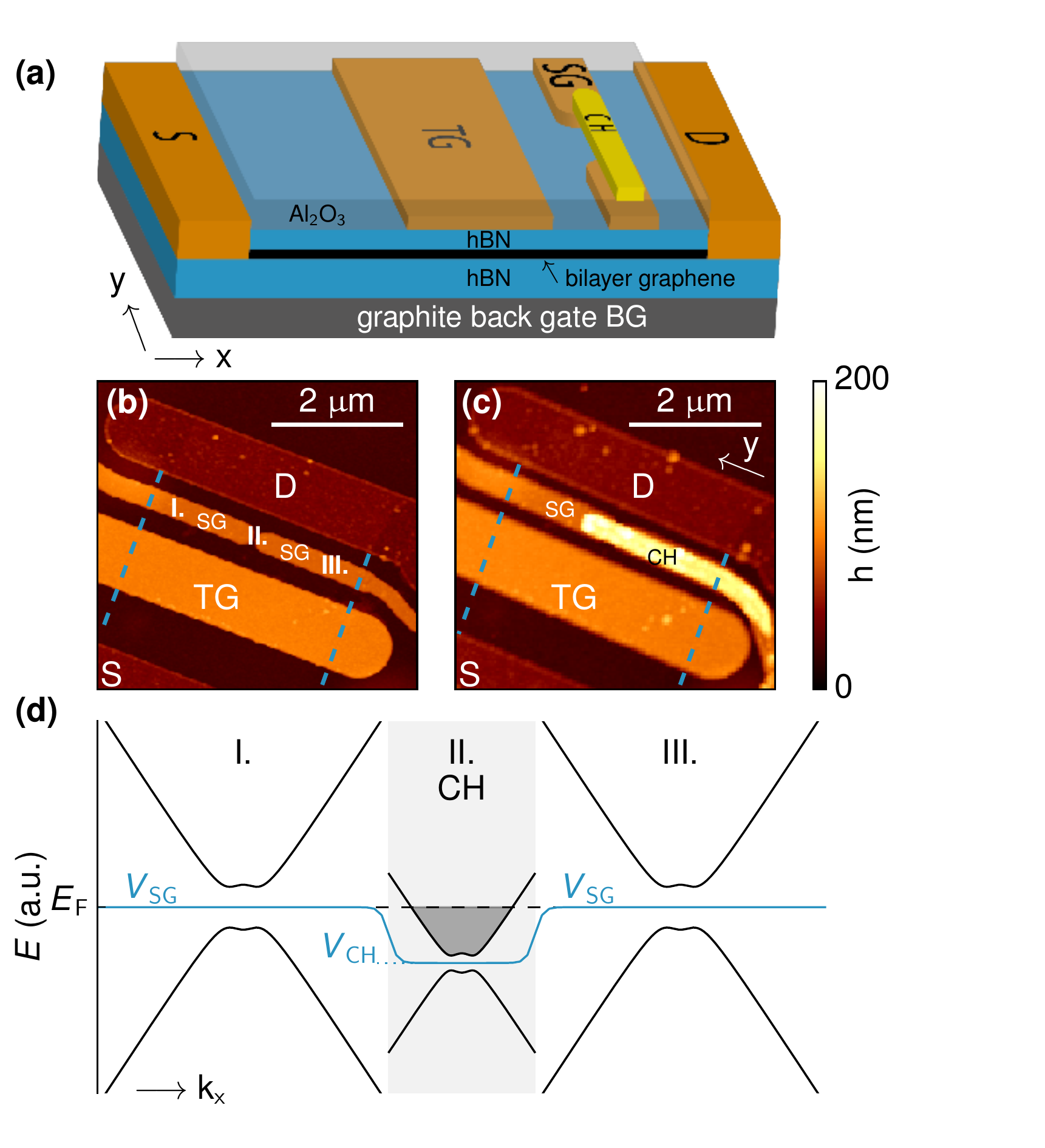}
\caption{ Sample layout. (a) Schematic of sample $A$. A bilayer graphene flake is encapsulated in hexagonal boron nitride. It is contacted by a source (S) and drain (D) contact and has a graphite back gate (BG) below, a top gate (TG), two split gates (SG) and a channel gate (CH) on top. The channel gate is separated from the split gates by a dielectric layer of Al$_2$O$_3$. (b) Atomic force microscopy image of the sample prior to deposition of the channel gate. The position of the graphene flake is indicated by blue dashed lines. (c) Atomic force microscopy image of the sample with the channel gate. (d) Model of the band structure along the y-direction with the electrostatic potential indicated by the blue line. The Fermi level under the split gates lies in the band gap. The channel gate induces a finite carrier density in the channel.}
\label{fig:1}
\end{figure}

Sample $A$, drawn schematically in Fig.~\ref{fig:1}a, consists of a stack of bilayer graphene encapsulated in hexagonal boron nitride on top of a graphite back gate. The stack was assembled using the van der Waals pick-up technique \cite{Wang2013} and was deposited on a Si/SiO$_2$ substrate chip. The probed graphene area is delimited by the two Ohmic contacts and the natural edges of the graphene flake (dashed blue lines in Fig.~\ref{fig:1}b,c). On top of the device, a  $1~\mu$m wide top gate (TG) and two $300~$nm wide split gates (SG), separated by $100~$nm, were evaporated (see atomic force microscopy image in Fig.~\ref{fig:1}b). Atomic layer deposition was performed to add a dielectric layer (Al$_2$O$_3$, $60~$nm). Finally, another $200~$nm wide gate, referred to as channel gate (CH), was evaporated onto the channel defined by the split gates (see Fig.~\ref{fig:1}c). Sample $B$ was produced in the same way but has a thinner Al$_2$O$_3$ layer ($20~$nm), a smaller channel width ($80~$nm) and a narrower channel gate ($60~$nm). In sample $B$, two separate pairs of contacts are used to probe either the graphene region with top gate, or the graphene region with split gate geometry. More details about the sample fabrication and geometry can be found in the Supplemental Material.

Unless stated otherwise the measurements were performed at $T = 1.7~$K. An AC bias voltage of $50~\mu$V was applied and the current $I$ was measured using low-frequency lock-in techniques.

In order to illustrate the basic idea of electrostatic confinement in bilayer graphene, we take a look at Fig.~\ref{fig:1}d. It shows a schematic of the $E(k)$ dispersion relation at three different points  across the  quantum point contact (QPC), indicated in Fig.~\ref{fig:1}b. When the Fermi level under the split gates lies in the gap (I., III.) and the Fermi level in the channel lies in the conduction band (II.), charge carriers can only flow through the narrow channel. A finite element simulation of the electrostatic potential can be found in the Supplemental Material.

\begin{figure}
\centering
\includegraphics[width=0.5\textwidth]{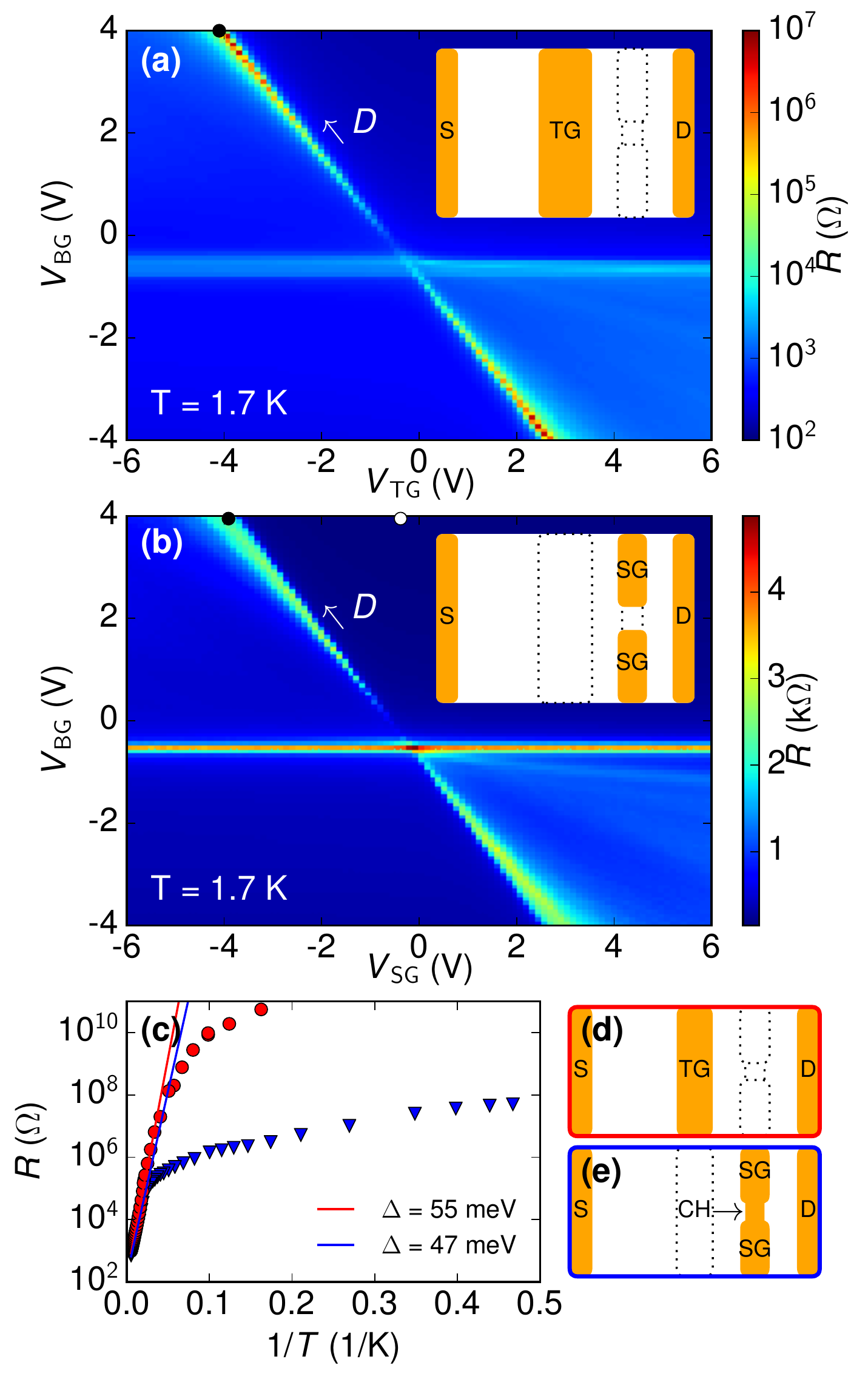}
\caption{Sample characterization of sample $A$. (a) Two-terminal resistance $R$ as a function of top gate voltage $V_\mathrm{TG}$ and back gate voltage $V_\mathrm{BG}$. The split gates and the channel gate were grounded. The diagonal line corresponds to charge neutrality underneath the top gate. Along this line the displacement field $D$ increases, which results in an increase of resistance. (b) Two-terminal resistance $R$ as a function  of split gate voltage $V_\mathrm{SG}$ and back gate voltage $V_\mathrm{BG}$. The channel gate and the top gate were grounded. In contrast to (a), the resistance does not increase with increasing displacement field, because charge carriers can flow through the channel. (c) Resistance $R$ as a function of temperature $T$ for the resistance maximum induced by the top gate and back gate (black dot in (a), schematic in (d)) and the resistance maximum induced by the combination of the split gates, the back gate (black dot in (b)) and the channel gate  at $V_\mathrm{CH} = -12$~V, schematic in (e)). Gap sizes of $\Delta = 55$~meV and $\Delta = 47$~meV were extracted from the high temperature behavior.}
\label{fig:2}
\end{figure}

To demonstrate experimentally that a band gap opens, we first look into the combined effect of the top gate (TG) and the back gate (BG), whilst keeping the split gate and the channel gate grounded. Figure \ref{fig:2}a shows the resistance of sample $A$ as a function of top gate voltage $V_\mathrm{TG}$ and back gate voltage $V_\mathrm{BG}$. The horizontal resistance maximum corresponds to the charge neutrality point of the outer regions of the sample which are not affected by the top gate voltage. The diagonal resistance maximum is the charge neutrality point of the sample region underneath the top gate. The displacement field $D$ increases in the direction of the arrow in Fig.~\ref{fig:2}a. It opens a band gap and hence increases the resistance at charge neutrality by several orders of magnitude. The global resistance maximum, indicated by a black dot in Fig.~\ref{fig:2}a, coincides with the point of highest displacement field $D = 0.7~$V/nm. The red dots in Fig.~\ref{fig:2}c show the evolution of the resistance maximum as a function of temperature (the corresponding configuration is sketched in Fig.~\ref{fig:2}d). Down to $T = 20~$K, the resistance follows an Arrhenius law ($R\sim \exp(\Delta/(2k_\mathrm{B}T)$) with a gap size $\Delta = 55~$meV. Below $T = 20~$K the resistance shows sub-exponential behavior, presumably because of hopping transport via mid-gap states. The resistance keeps increasing nonetheless. In this highly resistive regime, the resistance has been determined from the slope of $I-V$ traces with a DC bias voltage range of $V_\mathrm{DC} = \pm 10~$mV. At $T = 5 
~$K the resistance is $R \sim 10~$G$\Omega$, which is the maximum resistance measurable in our set-up. For sample $B$, we measure a maximum resistance of $R = 10~$M$\Omega$. In a third sample with a graphite back gate and a uniform top gate, we also measured resistances on the order of $R \sim 10~$G$\Omega$.

Resistances on the order of gigaohms are rarely observed in bilayer graphene\cite{Li2016a,Sui2015}. In most samples, a saturation of the resistance occurs in the megaohm range \cite{Weitz2010a,Taychatanapat2010,Xia2010} or below \cite{Zou2010,Yan2010,Zhu2016a}. Zibrov et al.\cite{Zibrov2017} already pointed out that the use of graphite gates can significantly reduce sample disorder. A high device quality with graphite gates was also reported in Ref.~\citenum{Li2017a}. The high resistance achieved in our three samples with a graphite back gate might be due to reduced disorder achieved by a better screening of charged impurities in the Si substrate, in the boron nitride and in the graphene itself, which leads to a reduction of the number of mid-gap states. The different stray field pattern arising from a close by back gate might also play a role, as it modifies the doping profile across the sample. The lower resistance maximum measured in sample $B$ compared to the other samples can be explained by the fact that the graphene region below the top gate showed some bubbles in the AFM image. The lower quality of the graphene in this region can lead to more mid-gap states. In any case, the resistance of $R = 10~$M$\Omega$ measured in this region is still significantly higher than the resistance quantum.

Figure~\ref{fig:2}b shows the resistance of sample $A$ as a function of split gate voltage (V$_\mathrm{SG}$) and back gate voltage (V$_\mathrm{BG}$), with a grounded top gate and channel gate. Lines of enhanced resistance follow the same pattern as in Fig.~\ref{fig:2}a. In contrast to Fig.~\ref{fig:2}a, the resistance along the displacement field axis does not increase beyond about $R = 5~$k$\Omega$ (note the different color scales of Figs.~\ref{fig:2})a and \ref{fig:2}b). This is because charge carriers can flow through the channel between the split gates. 

The channel can be depleted, however, by applying a channel gate voltage $V_\mathrm{CH} = -12~$V. The blue triangles in Fig.~\ref{fig:2}c show the resistance as a function of temperature for (V$_\mathrm{SG}$,V$_\mathrm{BG}) = (-3.9,4)~$V (black dot in Fig.~\ref{fig:2}b) and $V_\mathrm{CH} = -12~$V, which gives the highest resistance achievable at $T = 1.7~$K using the split gates and the channel gate (configuration in Fig.~\ref{fig:2}e). In the high temperature regime a gap energy of $\Delta = 47~$meV can be extracted. The resistance deviates from the activated behavior below $T \sim 50~$K and goes up to $R = 50~$M$\Omega$ at $T = 1.7~$K, which is three orders of magnitude higher than the resistance quantum $h/e^2$. In sample $B$ the maximal resistance achieved with the split gates and the channel gate is $R = 20~$M$\Omega$ at $T = 1.7~$K. These results are in contrast with previous works on bilayer graphene QPCs, which showed a minimal conductance above $G = e^2/h$.\cite{Allen2012,Goossens2012a} They show that it is not only possible to achieve high resistances with a rather wide uniform gate, but also with a combination of three narrower gates.  The band gap underneath the split gates is sufficient to suppress conductance when the Fermi energy is in the gap. We will therefore focus below on the conductance of the channel.

\begin{figure*}
\centering
\includegraphics[width=1\textwidth]{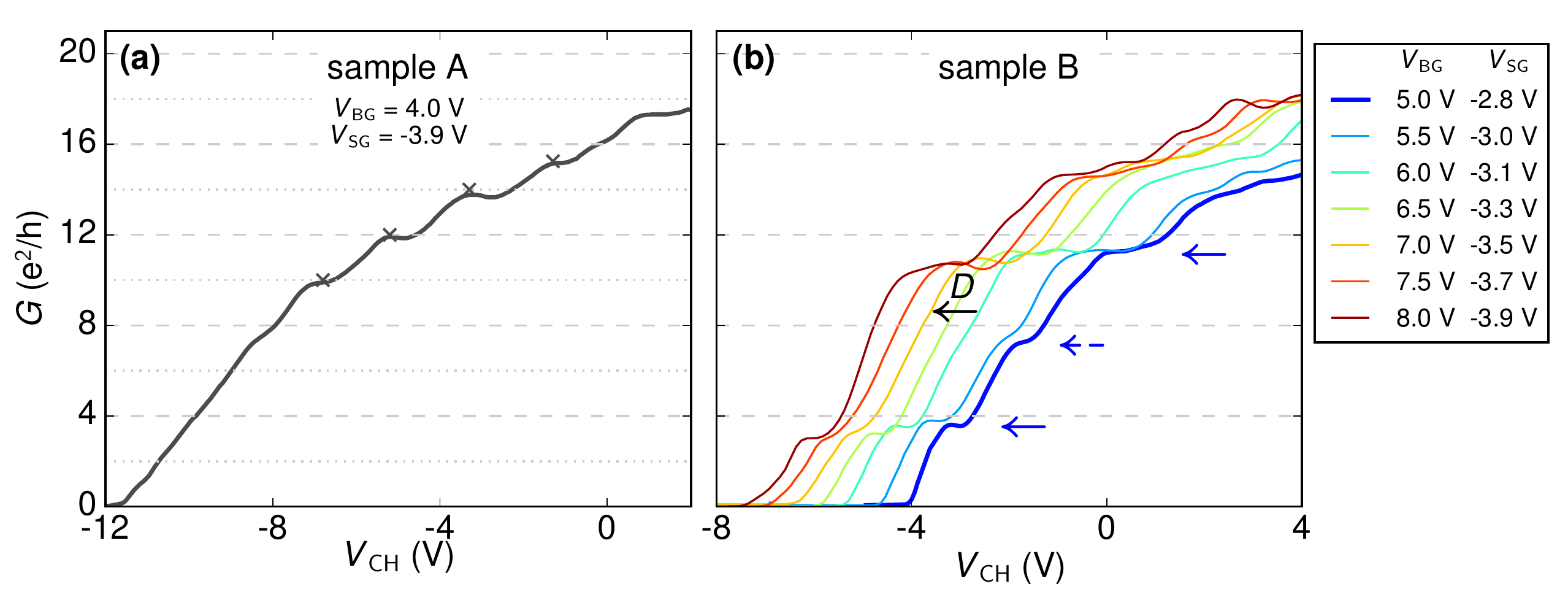}
\caption{(a) Conductance $G$ of the induced channel in sample $A$ as a function of  $V_\mathrm{CH}$ at $B = 0$~T for the gate voltage configuration at the black dot in Fig.~\ref{fig:2}b. The conductance shows a number of steps of $\Delta G = 2~e^2/h$.  (b) Conductance $G$ of the channel in sample $B$ as a function of channel gate voltage $V_\mathrm{CH}$ at $B = 0$~T for several combinations of back gate and split gate voltage. The conductance shows plateaus slightly below $G = 4,8,12~e^2/h$.}
\label{fig:3}
\end{figure*}

We vary the channel gate voltage $V_\mathrm{CH}$ in the regime where conductance under the split gates is maximally suppressed.   For sample $A$ the back gate voltage could not be increased above V$_\mathrm{BG} = 4~$V because of the onset of gate leakage, most likely due to the thin hBN layer between the back gate and the contacts. Suppression of conductance under the gates was only reached at (V$_\mathrm{SG}$,V$_\mathrm{BG}) = (-3.9,4)~$V (see black dot in Fig.~\ref{fig:2}b). The conductance $G$ at this operating point as a function of channel gate voltage $V_\mathrm{CH}$ is shown in Fig.~\ref{fig:3}a. A series resistance of $R_\mathrm{S} = 150~\Omega$ was subtracted, which was determined by measuring the resistance at (V$_\mathrm{SG}$,V$_\mathrm{BG}) = (-0.4,4)~$V (see white dot in Fig.~\ref{fig:2}b). This point corresponds to uniform doping throughout the sample. The conductance shows plateaus at $G = 8,10,12,14,16,18~e^2/h$. No plateaus are discernable below $G = 8~e^2/h$. To our knowledge this is the largest number of conductance plateaus observed in bilayer graphene to date. At $V_\mathrm{CH} = -12~$V the channel is depleted, reaching a resistance of $R = 50~$M$\Omega$.

Sample $B$ has a larger back gate voltage range with gate leakage smaller than 0.1 nA. Figure \ref{fig:3}b shows its conductance as a function of channel gate voltage for a set of back gate - split gate voltage pairs. Under the split gates, increasing voltage differences $V_\mathrm{BG}-V_\mathrm{SG}$ correspond to an increasing displacement field $D$ along the charge neutrality line (cf. Fig.~\ref{fig:2}b). For each curve, a series resistance equal to the resistance measured at uniform doping at the corresponding back gate voltage was subtracted. Throughout the whole range, plateaus can be observed slightly below $G = 4,12~e^2/h$ (see blue arrows). For $V_\mathrm{BG} < 6~$V a plateau occurs slightly below $8~e^2/h$ as well (dashed blue arrow). In the range above $G = 12~e^2/h$ small oscillations are observed which cannot be identified to be quantized conductance plateaus. 

Sample $B$ could also be depleted completely at $V_\mathrm{BG} = -8~$V, when the entire sample is $p$-doped. The conductance as a function of channel gate voltage in this setting shows several smaller kinks, but no quantized conductance plateaus. In sample $A$ it was not possible to deplete the channel in the $p$-doped regime, most likely because of the limited back gate voltage range.

Our results fit well into the landscape of experiments on single- and bilayer graphene QPCs published previously, where lifted degeneracies were observed in some but not all samples. Theoretically, for pristine bilayer graphene, steps of $\Delta G = 4~e^2/h$ are expected because of spin- and valley-degeneracy, as observed in sample $B$. However, the observed step size of $\Delta G = 2~e^2/h$ in sample $A$, witnessing a lifted degeneracy, is in agreement with other experimental works on bilayer graphene.\cite{Allen2012,Goossens2012a}  In monolayer graphene, conductance quantization with steps of $\Delta G = 2~e^2/h$ was observed in both limits of low\cite{Tombros2011,Somanchi2017a} and high\cite{Terres2016a} mode number. However, Kim et al.\cite{Kim2016a} reported conductance quantization with a step size $\Delta G = 4~e^2/h$ in an electrostatically induced channel in monolayer graphene. Zimmermann et al.\cite{Zimmermann2016a} studied a QPC in single layer graphene in the quantum Hall regime where a step size of $\Delta G = 1~e^2/h$ is observed.

We speculate that the difference in the observed degeneracies in samples $A$ and $B$ is caused by the residual disorder in these devices. In the quantum Hall regime all degeneracies in the lowest Landau level are lifted in our samples (see below), which demonstrates the good sample quality. Yet at zero magnetic field, the lack of perfect flatness of plateaus, the deviations from the expected plateau values, the occasionally missing plateaus, and the absence of plateaus in a $p$-doped channel indicate that a further increase in device quality, currently out of reach, would lead to improvements. Beyond that, strain effects could modify the potential landscape. In GaAs heterostructures, it is well known that a difference in thermal expansion coefficient between the metal gates evaporated on top of the semiconductor wafer and the semiconductor material itself can lead to a strain-induced change of the potential of 5-10\%.\cite{Ye1995} In our case, the hBN layer separating the metal gate from the graphene layer is comparatively thinner, and one can imagine that strain effects could also lead to modifications of the potential, in addition to the electrostatic definition of the QPC. While further improvements in device quality will lead to better reproducibility among different devices and allow for investigating more subtle interaction effects, such as spontaneous spin polarization\cite{Castro2008}, at present the microscopic origin of the lifted degeneracy and the missing plateaus at low mode numbers in device $A$ remains unknown.

\begin{figure*}
\centering
\includegraphics[width=1\textwidth]{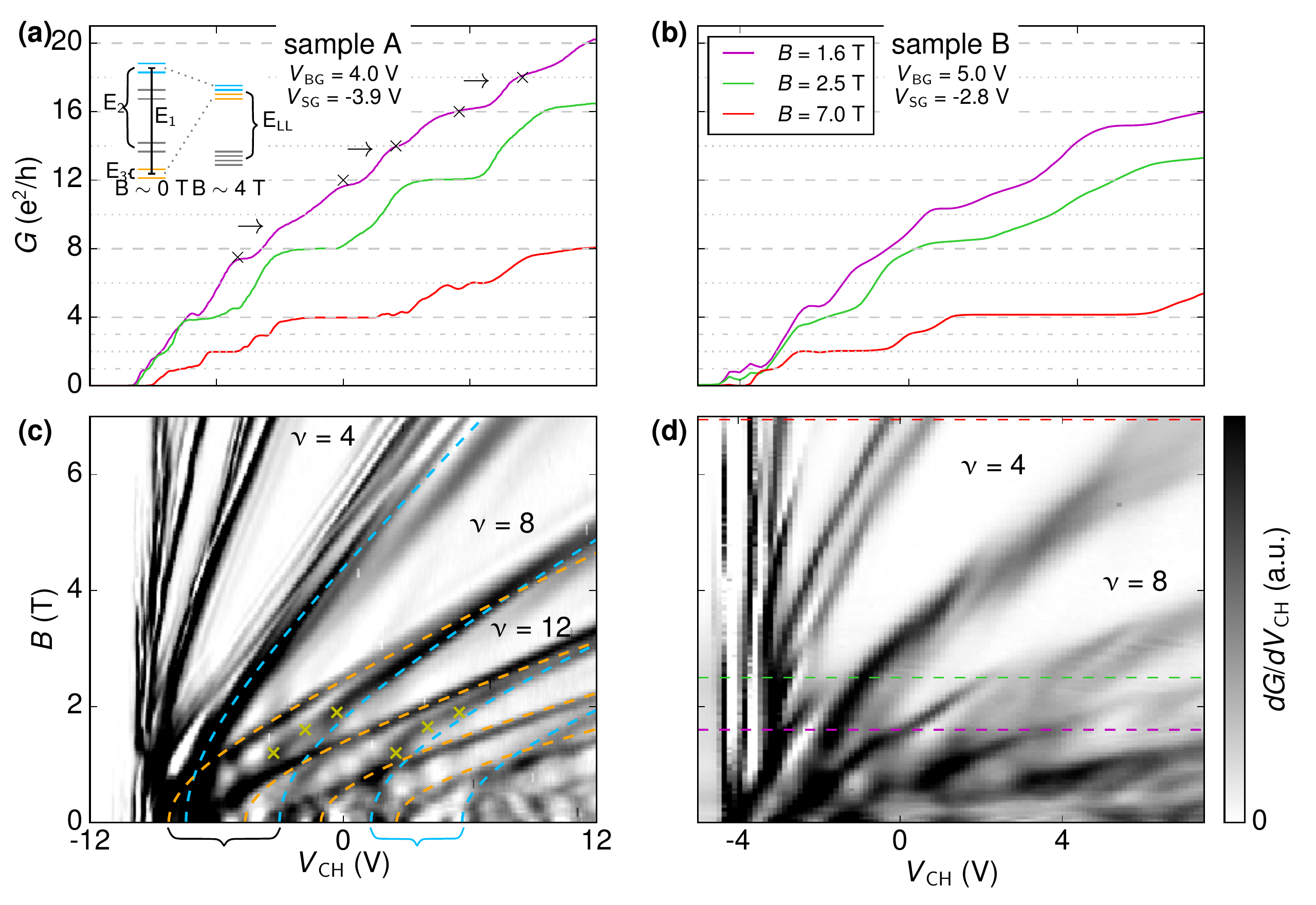}
\caption{(a) Conductance of sample $A$ for several magnetic field strengths. The plateaus at 10, 14 and 18 $e^2/h$ are still present at $B = 1.6$~T (see arrows), but disappear in higher magnetic fields. At $B = 7$~T (red line) plateaus are present at 1, 2 ,3, 4 $e^2/h$. (b) Conductance of sample $B$ for several magnetic field strengths. (c) Transconductance of sample $A$ as a function of channel gate voltage $V_\mathrm{CH}$ and magnetic field $B$. The blue and orange dashed lines both follow the model described by Eq.~\ref{eq:modes}. (d) Transconductance of sample $B$ as a function of channel gate voltage $V_\mathrm{CH}$ and magnetic field $B$. The transitions between modes are less pronounced than in sample $A$. Horizontal dashed lines correspond to the line cuts in (a),(b). }
\label{fig:4}
\end{figure*}

A magnetic field has the potential to give further insights into degeneracy lifting in QPCs. Figure \ref{fig:4}a(b) shows the conductance of sample $A$($B$) as a function of $V_\mathrm{CH}$ for selected magnetic field strengths. For these measurements, the density in the bulk of the sample is considerably higher than the density in the channel. The conductance is therefore governed by the filling factor of the channel (see Supplemental Material). In a magnetic field of $B = 7~$T we observe that the four-fold degeneracy of the lowest Landau level is completely lifted in both samples, demonstrating the high quality of the samples.\cite{Varlet2014a} Sample $A$ shows a step size of $\Delta G = 4~e^2/h$ at intermediate magnetic fields (see curve at $B = 2.5~$T). This is surprising, since the step size at $B = 0~$T (Fig.~\ref{fig:3}a) and $B = 1.6~$T (see arrows in Fig.~\ref{fig:4}a) is only $\Delta G = 2~e^2/h$ for this sample. In sample $B$, no clear quantization of the levels is observed at intermediate magnetic fields.

The transconductance as a function of $V_{CH}$ and $B$, shown in Fig.~\ref{fig:4}c,d, provides a more complete picture. Transitions between quantized modes are seen as dark lines. In sample $A$, the transitions between the plateaus are more pronounced than in sample $B$. In both samples, these lines start out vertically at low magnetic fields, and bend over between $B = 1~$T and $B = 2~$T towards more positive gate voltages, ending up as straight lines with finite slope at high fields. This behavior is reminiscent of the magnetoconductance of high quality QPCs, for example in GaAs, where the low magnetic field conductance is confinement dominated, whereas the high magnetic field conductance is determined by edge channels formed in crossed electric and magnetic fields. The effect is known as the magnetic depopulation of magnetoelectric subbands.\cite{Beenakker1991} Also in our samples, filling factors can be assigned to the light regions between the lines as indicated in the figure. However, when the magnetic field is decreased towards the confinement dominated regime, the mode structure appears to be much more complicated than in GaAs.

\begin{figure}
\centering
\includegraphics[width=0.5\textwidth]{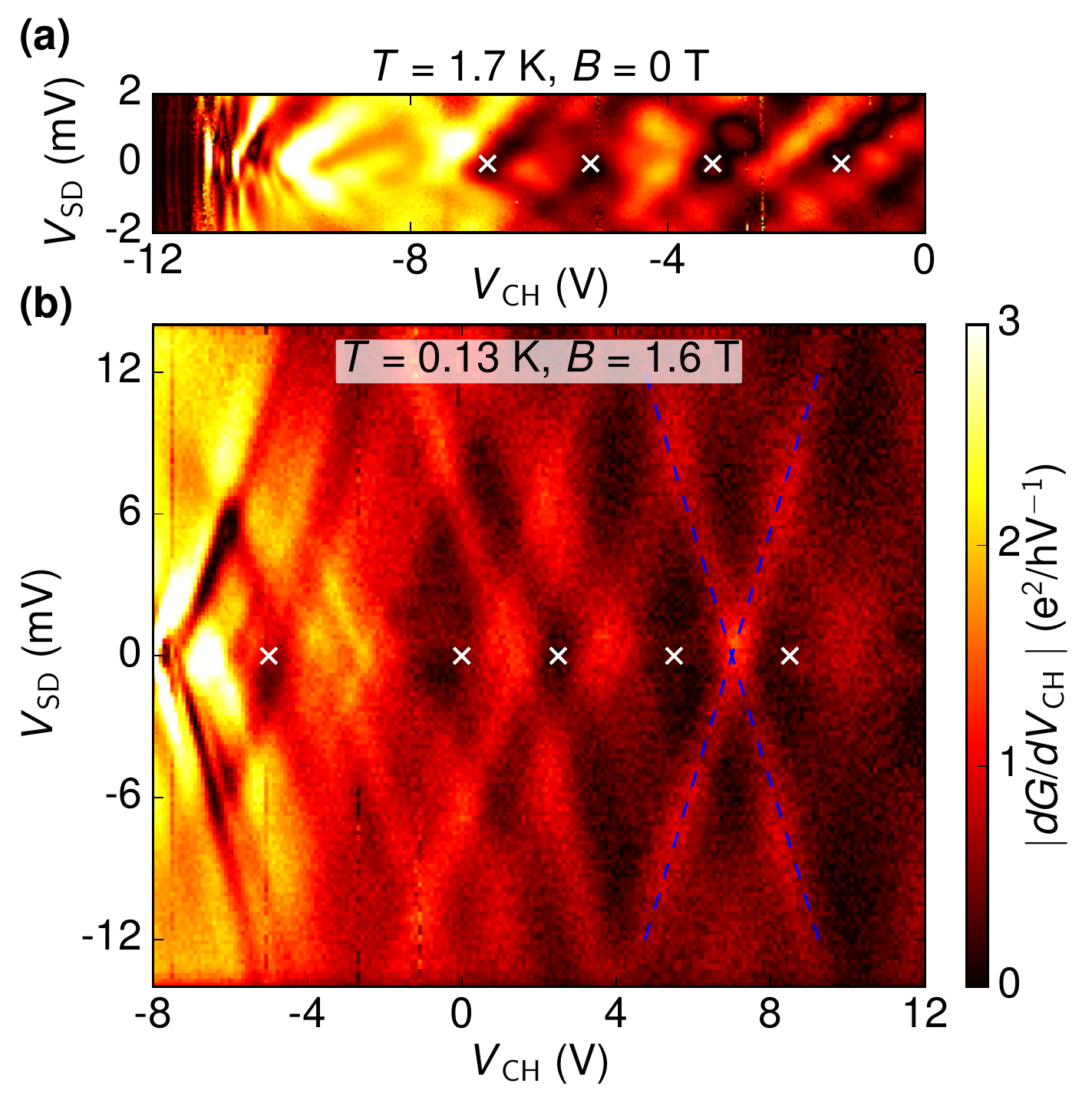}
\caption{(a) Transconductance $dG/dV_\mathrm{CH}$ of sample $A$ as a function of channel gate voltage $V_\mathrm{CH}$ and source drain bias $V_\mathrm{SD}$ for the gate configuration of the black dot in Fig.\ref{fig:2}b. The measurement was performed at $T = 1.7$~K and $B = 0$~T. At the positions of the white crosses, corresponding to the plateaus marked by crosses in Fig.~\ref{fig:3}a, minima in transconductance can be observed. (b) Transconductance at $B = 1.6$~T. The white crosses indicate the positions of the plateaus in Fig.~\ref{fig:4}a and coincide with the diamond shaped features observed here around $V_\mathrm{SD} = 0$~mV, even though the two measurements were recorded during different cooldowns.}
\label{fig:5}
\end{figure}

Lacking a detailed theory we propose a heuristic model which describes the level transitions of sample $A$ as a function of magnetic field. In analogy to magnetic depopulation in GaAs 2DEGs\cite{Buttiker1990}, we assume that the energy separation of the modes in the channel is given by
\begin{equation}
E_N = \hbar \Omega \sqrt{N(N-1)}, \qquad \Omega = \sqrt{\omega_0^2+\omega_c^2}
\label{eq:modes}
\end{equation}
where $\omega_0$ is the frequency of the electrostatic confinement potential in the absence of a magnetic field, and $\omega_c$ is the cyclotron frequency, given by $\omega_c = eB/m^*$. Assuming a linear conversion from gate voltage $V_\mathrm{CH}$ to energy $E = \alpha e
 (V_\mathrm{CH}-V)$, it is impossible to fit a mode spectrum as that described by Eq.~\ref{eq:modes} to all the levels observed in Fig.~\ref{fig:3}d using $\alpha,~V$ and $\omega_0$ as free fitting parameters. Yet by extending the model with a second set of parameters $\alpha',~V'$ and $\omega_0'$ it is possible to capture the trends of the level crossings in the low magnetic field regime. This is demonstrated by the dashed orange and blue lines in Fig.~\ref{fig:3}d. The employed parameters are $\hbar\omega_0 = 7.5~$meV, $\alpha = 1.75 \times 10^{-3}$, $V= 13.5~$V, $\hbar\omega_0' = 5~$meV, $\alpha' = 1.4 \times 10^{-3}$ and $V'= 17~$V. We want to stress that the model is purely heuristic. It was designed to capture the dominant features of the experiment. The two different frequencies could imply that the two valley/spin split modes may have different effective masses. The difference between $V$ and $V'$ indicates an energy offset between the two sets of levels. The model captures the main features of the data, except for the part where $V_\mathrm{CH} <-10~$V (where the conductance at $B = 0~$T already deviates from the expected pattern), and the features marked by yellow crosses in Fig.~\ref{fig:3}d. The parameters $\hbar \omega_0$ and  $\hbar \omega_0'$ are similar to the curvature of the harmonic potential calculated in a COMSOL simulation of the electrostatic potential of the device (see Supplemental Material). The parameters $\alpha$ and $\alpha'$ are in rough agreement with the slope of the finite bias diamond boundaries, which yield a lever arm of $2.6 \times 10^{-3}$ (see blue dashed lines in Fig.~\ref{fig:5}b.)

The data suggest that around $B = 4~$T, the spin and valley splittings are too small to be resolved. The only relevant energy spacing is the Landau level spacing $E_\mathrm{LL}$ (see inset of Fig.~\ref{fig:3}a). Lowering the magnetic field, the relative influence of the electrostatic potential compared to the magnetic confinement grows, which lifts a degeneracy (the blue and orange dashed lines move apart). The black curly bracket in Fig.~\ref{fig:3}d indicates the energy range of the lifted degeneracy at $B = 0~$T ($E_\mathrm{1}$ in Fig.~\ref{fig:3}a), which seems to have grown larger than the mode spacing indicated by the blue curly bracket ($E_\mathrm{2}$ in Fig.~\ref{fig:3}a). The remaining twofold degeneracy implies that the energy scale $E_\mathrm{3} = 0$. Although the model suggests a degeneracy lifting larger than the mode spacing of the QPC, we currently do not know which mechanism could be responsible for such a drastic effect.

Another aspect which may contribute to the crossing mode pattern is the fact that the channel gate voltage changes the displacement field $D$ inside the channel. Bilayer graphene exhibits a valley splitting of the Landau levels which depends on the displacement field \cite{Zhang2011,Lee2014,Hunt2016a}. In the devices presented here, the charge carrier density and the displacement field in the channel cannot be varied independently, complicating a systematic study of the effect of the displacement field.

Finite bias measurements were performed to extract subband energy spacings. Figure \ref{fig:5}a shows the transconductance $|dG/dV_\mathrm{CH}|$ as a function of source drain bias measured at $T = 1.7$~K and $B = 0$~T. Minima in the transconductance are observed at the positions of the plateaus in Fig.~\ref{fig:3}a (see crosses), but there is no simple diamond pattern. The energy spacing seems to be on the order of $\Delta E \approx 1~$meV. In sample $B$, features with a similar energy spacing are observed in finite bias measurements. In a finite magnetic field, a more pronounced diamond pattern is recovered. This can be seen in the transconductance measurement of sample $A$ in Fig.~\ref{fig:5}b, recorded at $T = 0.13$~K and $B = 1.6$~T. The centers of the diamonds correspond well to the conductance plateaus in Fig.~\ref{fig:3}d, even though these measurements were performed during different cooldowns. Compared to the level spacing at $B = 0~$T, the level spacing $\Delta E \approx 7~$meV at $B =1.6~$T shows a better agreement with the level spacing extracted from the heuristic model, which predicts a level spacing of $\Delta E = 7.5$~meV at $B = 1.6$~T for the orange line set in Fig.~\ref{fig:4}c.

In conclusion, we have demonstrated that bilayer graphene samples with graphite back gate show resistances of at least $R = 10~$M$\Omega$ in high displacement fields. This is a significant improvement compared to devices without a graphite back gate, where the resistance usually saturates in the regime of $R = 10 - 100~$k$\Omega$. \cite{Zou2010,Yan2010,Zhu2016a} We exploit this result to electrostatically define QPCs in bilayer graphene and observe the following:

1. In both samples, the channels can be fully depleted by gating.  

2. Both samples show quantized conductance though with different values for degeneracies.

3. Both samples show the expected quantum Hall plateaus with 4 fold degeneracies at high fields and complete lifting of degeneracies for the lowest Landau levels.

4. Both samples show an intricate crossover regime between zero magnetic field and quantum Hall regime where level crossings and avoided crossings occur.

The different step sizes of $\Delta G = 2~e^2/h$ and $\Delta G = 4~e^2/h$ in the two samples might be due to a different disorder potential or different strain patterns. Several factors, such as the reduced transmission of the modes of sample $B$ and the absence of conductance quantization in the $p$-doped regime, indicate that mesoscopic details of the samples play an important role. Realizing one-dimensional nanostructures in bilayer graphene by electrostatic gating paves the way towards controllable quantum dots in bilayer graphene.

\section*{acknowledgements}
We thank Anastasia Varlet, Leonid Levitov, Kostya Novoselov and Mansour Shayegan for fruitful discussions. We acknowledge financial support from the European Graphene Flagship and the Swiss National Science Foundation via NCCR Quantum Science and Technology. Growth of hexagonal boron nitride crystals was supported by the
Elemental Strategy Initiative conducted by the MEXT, Japan and JSPS
KAKENHI Grant Numbers JP15K21722.

\bibliography{qpc}

\clearpage


\section*{Supplementary material}

# Supporting Information for
# Electrostatically induced quantum point contacts in bilayer graphene

Hiske Overweg,[1, *] Hannah Eggimann,[1] Xi Chen,[2] Sergey Slizovskiy,[2] Marius Eich,[1] Riccardo Pisoni,[1] Yongjin Lee,[1] Peter Rickhaus,[1] Kenji Watanabe,[3] Takashi Taniguchi,[3] Vladimir Fal'ko,[2] Thomas Ihn, and Klaus Ensslin

[1]Solid State Physics Laboratory, ETH Zürich, CH-8093 Zürich, Switzerland
[2]National Graphene Institute, University of Manchester, Manchester M13 9PL, UK[†]
[3]National Institute for Material Science, 1-1 Namiki, Tsukuba 305-0044, Japan

| | sample A | sample B |
|---|---|---|
| graphite thickness (nm) | 28 | 15 |
| bottom BN thickness (nm) | 38 | 53 |
| top BN thickness (nm) | 35 | 25 |
| $Al_2O_3$ thickness (nm) | 60 | 30 |
| SG height (nm) | 60 | 20 |
| channel width (nm) | 100 | 80 |
| channel gate width (nm) | 200 | 60 |
| mobility (cm$^2$/Vs) | $8 \times 10^4$ | $6 \times 10^4$ |

TABLE I. Characteristics of samples A and B

## FABRICATION AND CHARACTERIZATION

The geometry and mobility of both samples can be found in Table I.

For the etching of the contacts, we use a recipe adapted from Ref. 1. We use a reactive ion etcher (Oxford Instruments RIE 80 Plus), with a mixture of CHF$_3$ gas (40 sccm) and O$_2$ (4 sccm). With an RF power of 60 W, the obtained etch rate of hBN is 45 nm/min. We carefully choose an etching time for each individual sample to make sure that the hBN is etched sufficiently for the contacts to reach the graphene layer, but not too far, since this would lead to a short between the contacts and the graphite back gate.

The deposition of Al$_2$O$_3$ was done in an atomic layer deposition system (Picosun Sunale R-150B) at a temperature of 150 ºC with trimethylaluminum (TMA) and water as precursor gases.

For sample A we performed temperature dependent measurements of the resistance maximum for (V$_{TG}$,V$_{BG}$) = -3.9, 4 V, corresponding to a displacement field of 0.7 V/nm, and (V$_{TG}$,V$_{BG}$) = -1.7, 2 V, corresponding to a displacement field of 0.4 V/nm. The respective gap sizes were 55 meV and 16 meV. An indirect measure of the gap size is the resistance at charge neutrality. Figure S1a shows the resistance of sample A along several line cuts in Fig. 2a. The resistance at the charge neutrality point increases by orders of magnitude when a band gap is opened. The measurement was performed with a bias voltage of V = 50 μV. The data for the highest displacement field (V$_{BG}$ = 4 V) has been omitted, because the high resistance peak cannot be reliably measured with a small bias

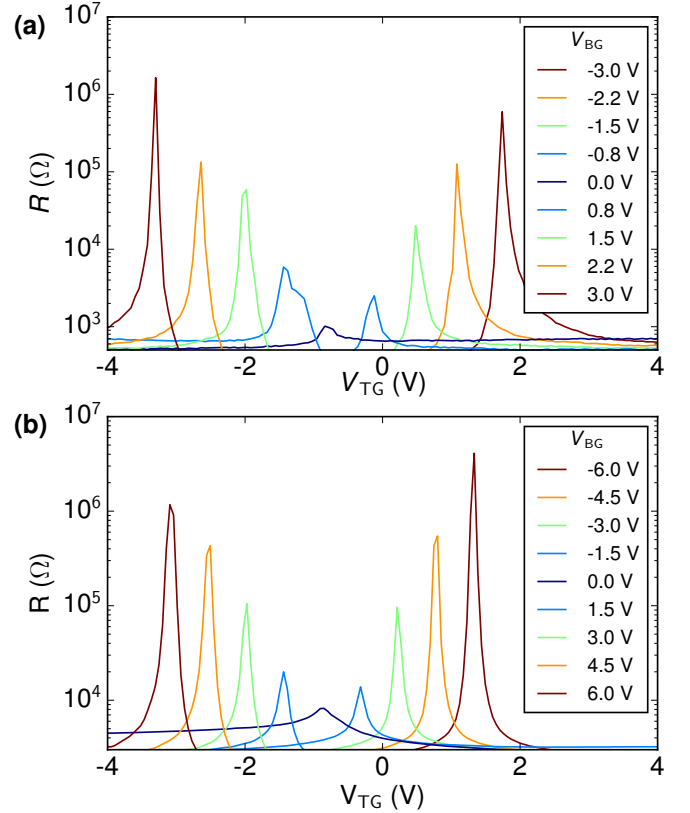

FIG. S1. (a) Resistance of sample A as a function of top gate voltage for several back gate voltages. The resistance at the charge neutrality point increases by orders of magnitude when a band gap is opened. (b) Same for sample B

voltage. Similar data for sample B is shown in Fig. S1b.

## SIMULATION OF THE ELECTROSTATIC POTENTIAL

To get more insight in the electrostatic potential of the quantum point contact, we use a finite element simulator (COMSOL). The charge carrier sheet density $n(\vec{r})$ and the potential $V(\vec{r})$ are calculated self-consistently using Poisson's equation and the Thomas-Fermi approximation. The density of states is approximated by $D(E) = m^*/(\pi \hbar^2) \theta(E)$ with $m^* = 0.034 m_e$ and $\theta$ the Heaviside step function, which limits the model to transport in the conduction band. We do not take the Mexi-

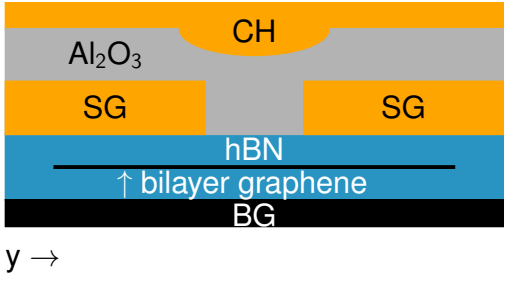

FIG. S2. Cross section of the sample used for Comsol simulation. The channel gate has an elliptic extension above the channel.

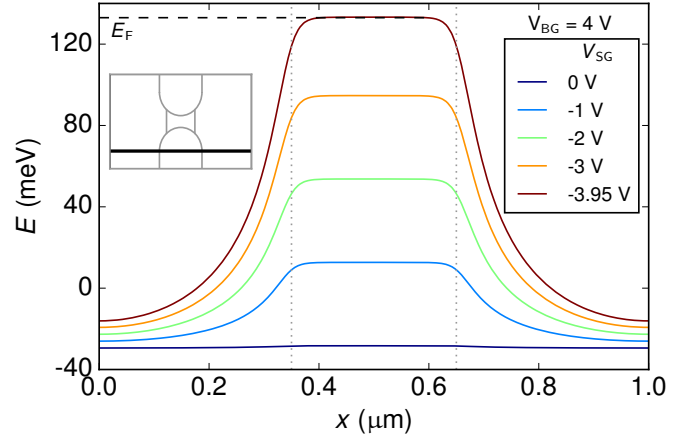

FIG. S3. Calculated electrostatic potential along a line cut (black line in inset) under the split gate for $V_{BG} = 4$ V and various split gate voltages. The dotted lines indicate the extent of the split gates. At $V_{SG} = 0$ V, the potential is constant throughout the structure. At $V_{SG} = -3.95$ V, the region under the split gates is depleted.

can hat shape of the bilayer graphene band structure and the position dependent band gap into account. The quantization of states inside the one-dimensional channel, due to the lateral confinement, is also neglected.

### Geometry

The geometry considered for the simulation is the same as for sample $A$ (see main text and table I). From AFM images of sample A it is apparent that the channel gate drops partially into the opening between the split gates. We therefore modelled the channel gate with an elliptical extension above the channel region (see Fig. S2). The width of this extension was 150 nm and the depth 30 nm, in agreement with the AFM images. When omitting the extension from the simulation, the channel gets depleted around $V_{CH} = -25$ V instead of the experimentally observed $V_{CH} = -12$ V. With the extension the channel gets depleted close to the experimentally observed value.

### Depletion below split gates

Figure S3 shows the calculated electrostatic potential along a line cut (black line in inset) under the split gate for $V_{BG} = 4$ V and various split gate voltages. At $V_{SG} = 0$ V, the potential is constant throughout the structure as expected. At $V_{SG} = -3.95$ V, the region under the split gates is depleted. This voltage is in agreement with the experimentally determined depletion voltage.

### Potential inside channel

In Fig. S4a the potential inside the channel is shown, for the case in which transport under the split gates is suppressed ($V_{BG} = 4$ V, $V_{SG} = -3.95$ V). For $V_{CH} = 0$ V, the potential can be approximated by a harmonic potential

$$V(y) = \frac{1}{2} m^* \omega_0^2 y^2$$

with an energy level separation of $\hbar\omega_0 = 8.4$ meV. For the gate voltage range of $V_{CH} = -10$ V - 12 V the energy level separation changes according to

$$\hbar\omega_0(V_{CH}) = 8.4 \text{ meV} + \alpha e V_{CH}$$

with $\alpha = 0.33 \times 10^{-3}$. The increased mode spacing with higher channel gate voltage can also be observed in Fig. 3: the conductance rises less steeply for higher channel gate voltage.

Figure S4b shows the potential across the QPC. For the range of $V_{CH} = -12$ V - 8 V the positive curvature in y-direction and the negative curvature in x-direction lead to a conventional saddle point potential. Similar results were obtained for simulations of sample $B$.

## ROLE OF THE BULK IN MAGNETOTRANSPORT

The density in the bulk of the sample is higher than the density inside the constriction for the entire range of Fig. 4. Only the edge modes that exist in both the bulk and the channel contribute to transport and the conductance is given by $G = \nu_{CH} e^2/h$. We can see a modest influence of the bulk of the device whenever the bulk is at a transition between integer filling factors. Figure S5 shows the derivative of the conductance with respect to magnetic field as a function of $V_{CH}$ and $B$ for the same gate voltage settings as Fig. 4. The horizontal lines, which occur in a $1/B$ periodic fashion, correspond to Landau level transitions in the bulk of the sample. The filling factors of the bulk are indicated on the y-axis. Because of the high charge carrier density, no broken degeneracies are observed up to $B = 8$ T.

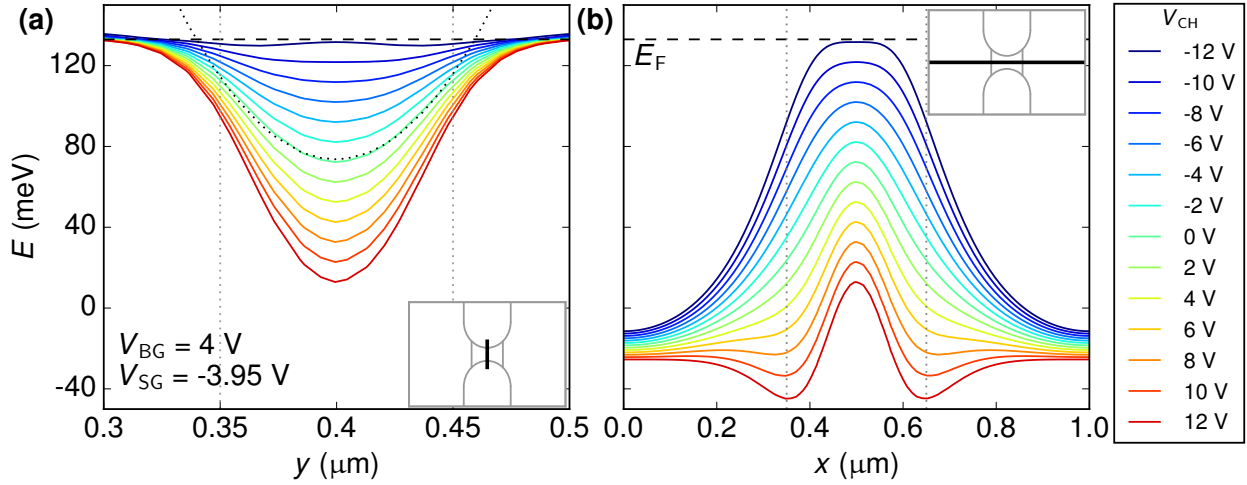

FIG. S4. Potential landscape for $V_{BG} = 4$ V, $V_{SG} = -3.95$ V and various channel gate voltages. (a) Electrostatic potential across the channel, which can be approximated by a harmonic potential (dotted black line). The channel gets depleted close to the experimental depletion voltage of $V_{CH} = -12$ V. (b) Electrostatic potential along the channel. For the range of $V_{CH} = -12$ V - 8 V a conventional saddle point potential is observed.

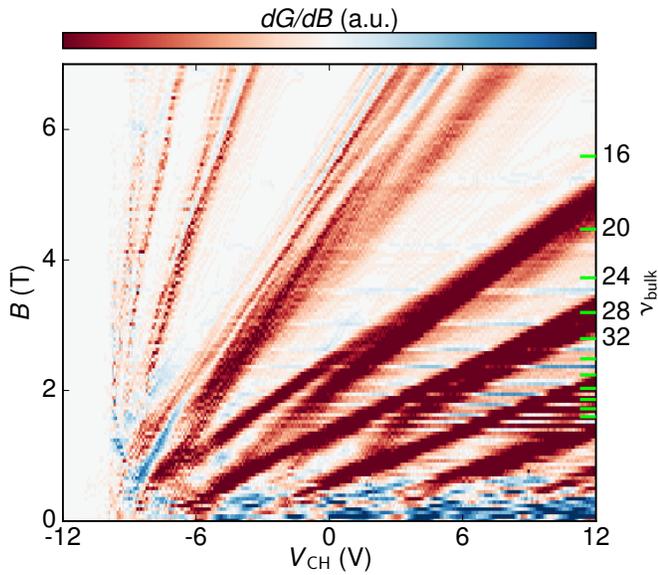

FIG. S5. Derivative of the conductance with respect to magnetic field for the same gate voltage settings as Fig. 4c. 1/B periodic horizontal lines are observed, which correspond to Landau level transitions in the bulk of the sample.

* overwegh@phys.ethz.ch
† On leave of absence from NRC "Kurchatov Institute" PNPI, Russia



\end{document}